\documentstyle[12pt]{article}
\input{tcilatex}
\begin{document}

\author{A. de Souza Dutra\thanks{%
E-mail: dutra@feg.unesp.br} \\
%EndAName
UNESP-Campus de Guaratinguet\'a-DFQ\\
Av. Dr. Ariberto Pereira da Cunha, 333\\
C.P. 205\\
12516-410 Guaratinguet\'{a} SP Brasil}
\title{{\LARGE Mapping deformed hyperbolic potentials into nondeformed ones}}
\maketitle

\begin{abstract}
In this work we introduce a mapping between the so called deformed
hyperbolic potentials, which are presenting a continuous interest in the
last few years, and the corresponding nondeformed ones. As a consequence, we
conclude that these deformed potentials do not pertain to a new class of
exactly solvable potentials, but to the same one of the corresponding
nondeformed ones. Notwithstanding, we can reinterpret this type of
deformation as a kind of symmetry of the nondeformed potentials.
\end{abstract}

\newpage

In this work we are interested in analyzing the so called deformed
hyperbolic potentials, as introduced by Arai \cite{arai} some years ago. We
intend to show that this interesting idea should be reinterpreted as a kind
of parameter scaling symmetry of the model. The deformed hyperbolic
functions introduced by Arai \cite{arai}, are given by 
\begin{equation}
\sinh _{q}\left( \alpha x\right) \equiv \frac{e^{x}-q\,e^{-x}}{2},\cosh
_{q}\left( \alpha \,x\right) \equiv \frac{e^{x}+q\,e^{-x}}{2}.
\end{equation}
In fact, we show here that by doing a convenient translation of the spatial
variable, one can transform the deformed potentials into the corresponding
nondeformed ones.

\begin{equation}
\sinh _{q}\left( \alpha x\right) =\sqrt{q}\,\sinh \left( \alpha \,y\right)
;\,\,\cosh _{q}\left( \alpha \,x\right) =\sqrt{q}\,\cosh \left( \alpha
\,y\right)
\end{equation}

\noindent where 
\begin{equation}
x=y+\frac{1}{\alpha }\ln \left( \sqrt{q}\right) .  \label{t1}
\end{equation}

As examples, we are going to explore some cases of a growing list of models
and works devoted to this kind of deformation \cite{arai}-\cite{egrifes4}.
Let us first examine the deformation of the Rosen-Morse potential, used
usually to treat molecular interactions, as studied by E\u {g}rifes et al 
\cite{egrifes}, and another one connected to it. They are given respectively
by 
\begin{equation}
V_{q}^{I}\left( x\right) =B_{0}\,\tanh _{q}\left( \alpha \,x\right)
-U_{0}\,\sec h_{q}^{2}\left( \alpha \,x\right) ,
\end{equation}

\noindent and 
\begin{equation}
V_{q}^{II}\left( x\right) =\frac{V_{1}}{2}\,\left( 1+\tanh _{q}\left( \alpha
\,x\right) \right) +\frac{V_{2}}{4}\,\left( 1+\tanh _{q}^{2}\left( \alpha
\,x\right) \right) .
\end{equation}

\noindent In fact, the first of the above potentials was recently considered
in a study of the Klein-Gordon equation \cite{egrifes4}.

It is easy to check that, by performing the above defined translation (\ref
{t1}), one recovers the corresponding nondeformed potentials: 
\begin{equation}
V_{q}^{I}\left( y\right) =B_{0}\,\tanh \left( \alpha \,y\right) -\frac{U_{0}%
}{q}\,\sec h^{2}\left( \alpha \,y\right) ,
\end{equation}

\noindent and 
\begin{equation}
V_{q}^{II}\left( y\right) =\frac{V_{1}}{2}\,\left( 1+\tanh \left( \alpha
\,y\right) \right) +\frac{V_{2}}{4}\,\left( 1+\tanh ^{2}\left( \alpha
\,y\right) \right) .
\end{equation}

For the first case, we see that the deformed system is nothing but a
translation with a corresponding parameter scaling of the well known
Rosen-Morse potential. This shows that there exists a symmetry connecting
the Rosen-Morse potential with different coupling parameters. In other
words, apart from a shift of the minimum of the potential, the energy
spectrum of the potential for different parameters is exactly the same.
Regarding the second potential examined in \cite{egrifes}, there is even no
change of the potential parameters, as can be observed in the figures
appearing in that paper.

By starting with the original form of the Rosen-Morse potential \cite
{rosenmorse}, 
\begin{equation}
V_{RS}\left( y\right) =B\tanh \left( \frac{y}{d}\right) -C\,\sec h^{2}\left( 
\frac{y}{d}\right) ,
\end{equation}

\noindent and identifying the parameters: $d=\frac{1}{\alpha }$. Then,
performing the translation $y=x-\frac{1}{\alpha }\ln \left( \sqrt{q}\right) $%
, we obtain the following transformed potential 
\begin{equation}
V_{RS}\left( x\right) =B\tanh \left( \alpha \,x\right) -\frac{C}{q}\,\sec
h^{2}\left( a\,x\right) \equiv V_{q}^{I}\left( y\right) ,
\end{equation}

\noindent so that, in order to complete the connection, we must have that 
\begin{equation}
B\equiv B_{0}\,,\,C\equiv U_{0}.
\end{equation}

In other words, we just need to change $B$ by $B_{0}$, $C$ by $\frac{U_{0}}{q%
}$ in the expression of the energy obtained in \cite{rosenmorse}. So we
obtain 
\begin{equation}
E_{n}=-\frac{1}{4}\left[ \left( 4\,\frac{U_{0}}{q}+g^{2}\right) ^{\frac{1}{2}%
}-g\,\left( 2\,n+1\right) \right] ^{2}+\frac{B_{0}^{2}}{\left[ \left( 4\,%
\frac{U_{0}}{q}+g^{2}\right) ^{\frac{1}{2}}-g\,\left( 2\,n+1\right) \right]
^{2}},
\end{equation}

\noindent where $g\equiv \frac{\hbar ^{2}\alpha ^{2}}{2\,M}$. From above, we
can see that changing the potential parameter $U_{0}$ and correspondingly
choosing a convenient modification of the translation parameter $q$, we
recover the same energy spectrum.

On the other hand, the corresponding eigenfunctions become 
\begin{eqnarray}
\psi _{n}\left( x\right) &=&N_{n}\,e^{-\,a\,\alpha \,x}\,\cosh ^{-b}\left(
\alpha \,x\right) \,\times  \nonumber \\
&&\times F\left( -n,\left( 4\gamma +1\right) ^{\frac{1}{2}}-n;a+b+1;\frac{1}{%
2}\left[ 1+\tanh \left( \alpha \,x\right) \right] \right) ,
\end{eqnarray}

\noindent where $N_{n}$ is the normalization constant which we will not
specify, once it is not necessary for the purpose of the present work.
Furthermore we have defined, 
\begin{equation}
a\equiv -\,g^{-2}\frac{B}{\left[ \left( 4\gamma +1\right) ^{\frac{1}{2}%
}-2\,n-1\right] };\,b\equiv \left( \gamma +\frac{1}{4}\right) ^{\frac{1}{2}%
}-\,n-\frac{1}{2};
\end{equation}

\noindent and $\gamma \equiv g^{-2}C$. Now, by doing our convenient
translation we recover: 
\begin{eqnarray}
\psi _{n}\left( y\right) &=&N_{n}\,q^{\frac{a-b}{2}}e^{-a\,\alpha
\,y}\,\cosh ^{-b}\left( \alpha \,y\right) \times  \nonumber \\
&&\times \,F\left( -n,\left( 4\gamma +1\right) ^{\frac{1}{2}}-n;a+b+1;\frac{1%
}{2}\left[ 1+\tanh \left( \alpha \,y\right) \right] \right) .
\end{eqnarray}

For the other hand, we can also discuss some other cases of the increasing
list of works discussing such kind of deformation. For instance, we can cite
two recent works \cite{jia3}, \cite{jia9}, which treat a five-parameter
exponential-type potential model, both in nonrelativistic \cite{jia3} as in
the relativistic Klein-Gordon case \cite{jia9}. In particular, they deal
with a superpotential with the following basic form 
\begin{equation}
W\left( x,Q_{1},Q_{2},Q_{3}\right) \equiv Q_{1}+\frac{Q_{2}}{e^{2\,\alpha
\,x}+q}+\frac{Q_{3}}{e^{2\,\alpha \,x}+q}e^{\alpha \,x},
\end{equation}

\noindent which, after performing the translation (\ref{t1}) suggested above
in the text, one ends with 
\begin{equation}
W\left( x,Q_{1},Q_{2},Q_{3}\right) \equiv Q_{1}+\frac{\left( \frac{Q_{2}}{q}%
\right) }{e^{2\,\alpha \,x}+1}+\frac{\left( \frac{Q_{3}}{\sqrt{q}}\right) }{%
e^{2\,\alpha \,x}+1}e^{\alpha \,x},
\end{equation}

\noindent where one can easily see that a scaling of the parameters
eliminates the dependence on the parameter $q$. So, we conclude that there
are only four independent parameters after the translation. Finally, let us
discuss the so called generalized Morse potential \cite{yelsitas}, given by 
\begin{equation}
V\left( x\right) =V_{1}e^{-2\,\alpha \,x}-V_{2}\,e^{-\alpha \,x}.
\end{equation}

\noindent In this case we perform the following translation 
\begin{equation}
x=y+\frac{1}{\alpha }\ln \left( \frac{V_{1}}{V_{2}}\right) ,
\end{equation}

\noindent which leads to 
\begin{equation}
V\left( x\right) =\left( \frac{V_{2}^{2}}{V_{1}}\right) \left( e^{-2\,\alpha
\,x}-e^{-\alpha \,x}\right) .
\end{equation}

From which we conclude that there exists just one effective potential
parameter, which implies into a symmetry of the parameters, leading to the
generation of a class of systems with different potential parameters but
with the same spectrum.

We could proceed further by discussing many systems presented in the
references below, with more or less details. Notwithstanding, we think that
the essential feature present in these potentials, as was illustrated
through the cases studied here, is that these so called deformed potentials
in fact pertain to the same class of potentials as the nondeformed ones.
They rather indicate a symmetry of the system where different potential
parameters correspond to the same spectrum and do not lead to new physical
insights.

\vfill 

\noindent {\bf Acknowledgments:} The author is grateful to CNPq for partial
financial support.

\end{document}